\documentclass{emulateapj}
\usepackage{apjfonts}

\shorttitle{Study of LS 5039 with {\it Suzaku}}
\shortauthors{Takahashi et al.}

\newcommand{\LS}{{LS~5039}}
\newcommand{\PSR}{{PSR~B1259$-$63}}
\newcommand{\LSI}{{LS~I~$+$61$\degr$303}}

\begin{document}

\title{Study of the Spectral and Temporal Characteristics of X-Ray Emission of the Gamma-Ray Binary LS~5039 with 
{\it Suzaku}}

\author{Tadayuki Takahashi\altaffilmark{1,2},
Tetsuichi Kishishita\altaffilmark{1,2},
Yasunobu Uchiyama\altaffilmark{3},
Takaaki Tanaka\altaffilmark{3},
Kazutaka Yamaoka\altaffilmark{4},
Dmitry Khangulyan\altaffilmark{5}
Felix  A. Aharonian\altaffilmark{5,6},
Valenti Bosch-Ramon\altaffilmark{5},
Jim A. Hinton\altaffilmark{7}
}
\email{takahasi@astro.isas.jaxa.jp}

\altaffiltext{1}{Institute of Space and Astronautical Science/JAXA, 3-1-1 Yoshinodai, Sagamihara, Kanagawa 229-8510, Japan}
\altaffiltext{2}{Department of Physics, University of Tokyo, 7-3-1 Hongo, Bunkyo, Tokyo, 113-0033, Japan}
\altaffiltext{3}{Kavli Institute for Particle Astrophysics and Cosmology, SLAC National Accelerator Laboratory, 2575 Sand Hill Road M/S 29, Menlo Park, CA 94025}
\altaffiltext{4}{Graduate School of Science and Engineering, Aoyama Gakuin University, 5-10-1 Fuchinobe, Sagamihara, Kanagawa 229-8558, Japan}
\altaffiltext{5}{Max-Planck-Institut f\"ur Kernphysik, Saupfercheckweg 1, Heidelberg 69117, Germany}
\altaffiltext{6}{Dublin Institute for Advanced Studies, 31 Fitzwilliam Place, Dublin 2, Ireland}
\altaffiltext{7}{School of Physics \& Astronomy, University of Leeds, LS2 9JT, Leeds, UK}

\begin{abstract}
We report on the results from {\it Suzaku} broadband X-ray observations  of the galactic binary source {\LS}. 
The {\it Suzaku} data, which have continuous coverage of more than one orbital period, show strong modulation of the
X-ray emission at the orbital period of this TeV gamma-ray emitting system.
The X-ray emission shows a minimum at orbital phase $\sim  0.1$, 
close to the so-called superior conjunction of the compact object,  
and a maximum at phase $\sim 0.7$, very close to the inferior conjunction of the 
compact object.  
The X-ray spectral data up to 70~keV are described by a hard power-law with a  phase-dependent photon 
index which varies within $\Gamma \simeq 1.45$-- 1.61. The amplitude of the flux variation is a factor of 2.5, 
but is significantly less than that of the factor $\sim$8 variation in the TeV flux. Otherwise the two light curves are similar, but not identical. 
Although  periodic  X-ray emission has been found from many  
galactic binary systems,  the {\it Suzaku} result implies a phenomenon  different from the 
``standard''  origin of X-rays related to the emission of the hot accretion plasma  formed 
around the compact companion object. 
The X-ray radiation of {\LS} is likely to be linked to very-high-energy electrons which are also responsible for the TeV gamma-ray emission. 
While the gamma-rays are the result of  inverse Compton scattering by electrons on optical stellar photons, 
X-rays  are produced via synchrotron radiation. Yet, while the modulation of the TeV gamma-ray signal can be naturally explained by the 
photon-photon pair production and anisotropic inverse Compton scattering, the observed modulation of 
synchrotron X-rays requires  an additional process, the most natural one being 
adiabatic expansion in the radiation production region. 
\end{abstract}


\section{Introduction}

{\LS} is a high-mass X-ray binary \citep{Motch97} with 
 extended radio emission \citep{Paredes00,Paredes02}.
This system is formed by a main sequence O type star and a compact object of 
disputed nature that has been claimed to be both 
a black hole \citep[e.g.,][]{casares05} and a neutron star/pulsar \citep[e.g.,][]{martocchia05,Dubus06}.
The compact object is moving around  the companion star in a moderately elliptic orbit (eccentricity $e=0.35$) 
with an orbital period of $P_{\rm orb} = 3.9060$ days \citep{casares05}.

As summarized in \cite{vbosch07}, LS~5039 has been observed
several times in the X-ray energy band for limited phases in the orbital period. Flux variations on time
scale of days and sometimes on much shorter timescales have been reported.
The spectrum was always well represented by a power-law model with a
photon index ranging $\Gamma=1.4$--1.6 up to $\sim 10$~keV, with fluxes
changing
moderately around $\sim 1\times 10^{-11}~{\rm erg}~{\rm cm}^{-2}~{\rm s}^{-1}$.
Softer spectra and larger fluxes had been also
inferred from {\it RXTE} observations, although background contamination
was probably behind these differences (see
\citealt{vbosch05}). Also, {\it Chandra} data taken in 2002 and 2005
showed spectra significantly harder than 1.5 \citep{vbosch07}, but
such a hard spectrum was probably an artifact produced by photon pile-up.
Recently, \cite{hoffmann} reported the
results of {\it INTEGRAL} observations in hard X-rays. The source was detected
at energies between 25 and 60~keV. The source was detected at energies between 25 and 60~keV. The flux was estimated
to be  $(3.54 \pm 2.30)  \times 10^{-11}~{\rm erg}~{\rm cm}^{-2}~{\rm s}^{-1}$ 
(90 \% confidence level) 
around the inferior conjunction (INFC) of the compact object, and a flux upper limit of
$1.45 \times 10^{-11}~{\rm erg}~{\rm cm}^{-2}~{\rm s}^{-1}$  (90 \% confidence level)  was
derived at the superior conjunction of the compact object (SUPC).

{\LS} has also been detected in very-high-energy (VHE: $E \geq 0.1$~TeV) gamma-rays  \citep{HESS_5039}, exhibiting a periodic 
signal modulated with the orbital period \citep{HESS_06}.  There are two other binary systems with robust detections in the
VHE range: {\PSR}  \citep{HESS_1259} and  {\LSI} \citep{MAGIC_LSI,VERITAS_LSI}. Evidence for TeV emission has been found also in
Cygnus~X-1 \citep{MAGIC_CygX1}. {\PSR} is a clear case of a high-mass binary system containing a non-accreting pulsar 
\citep{Johnston92}, whereas Cygnus X-1 is a well known accreting black hole system \citep{Bolton72}.  The nature of the
compact object in {\LS} is not yet established, and the origin of  the VHE emitting electrons is unclear.   They may be
related to a pulsar wind or to a black hole with a (sub)relativistic jet.   In the standard  pulsar-wind scenario,
the severe photon-photon absorption  makes the explanation of the detection  of the VHE radiation problematic, at least at the position
corresponding  to the orbital phase $\phi\sim 0.0$ (see Figs. 16 and 4 from
Sierpowska-Bartosik \&  Torres 2008 and Dubus et al. 2008 and compare with Fig. 5 in Aharonian et al. 2006), in which the
emitter is expected to be located between the compact object and the star.  However, particles may be accelerated in a
relativistic outflow formed at the interaction of the pulsar and  the stellar winds \citep{Bogovalov08}  and radiate far from
the compact object, making the pulsar wind scenario a viable option.  In the microquasar scenario, the lack of
accretion features in the X-ray spectrum may be a  problem unless the bulk of the accretion power is released in the form of
kinetic energy of the outflow,  rather than thermal emission during accretion, as in the case  of SS~433 \citep[see
e.g.,][]{Marshall02}.   At this stage, we cannot give a preference to any of these scenarios, but new data, in particular
those  obtained with the {\it Suzaku} satellite, allow us to make an important step towards the understanding of  the nature
of the non-thermal processes of acceleration and radiation in this mysterious object.

\section{Observation}

The temporal and spectral characteristics of the X-ray emission from {\LS} along the orbit 
should provide important clues for understanding the acceleration/radiation 
processes in this source. 
The fact that all previous X-ray observations of this object have incomplete coverage of the orbital period, 
or suffered from background contamination, is therefore rather unsatisfactory.
This motivated our long, $\sim 200$~ks observation with 
the {\it Suzaku} X-ray observatory \citep{mit07},  
which gives us unprecedented coverage
of more than one orbital period, continuously 
from 2007 September 9 to 15 (see Table \ref{tab:ls5039_obs}). 
{\it Suzaku} has four sets of X-ray telescopes \citep{ser07}
each with a focal-plane X-ray CCD camera \citep[X-ray Imaging Spectrometer(XIS);][]{koy07} 
that are sensitive in the energy range of 0.3--12~keV. 
Three of the XIS detectors  (XIS0, 2 and 3) have front-illuminated (FI) CCDs, 
whereas XIS1 utilizes a back-illuminated (BI) CCD. 
The merit of the BI CCD is its improved sensitivity in the soft X-ray energy band below 1~keV.
{\it Suzaku} contains also a non-imaging collimated Hard X-ray Detector \citep[HXD;][]{tak07,kok07}, 
which covers the 10--600~keV energy band with Si PIN photodiodes (10--70~keV) and GSO scintillation detectors 
(40--600~keV). 
{\it Suzaku} has two default pointing positions, the XIS nominal position and the HXD nominal position.
In this observation, we used the HXD nominal position, in which the effective area of  HXD
is maximized, whereas that of the XIS is reduced to on average $\sim$~88\%.
Results from XIS2 are not reported here since it has not been in operation since 
an anomaly in November 2006. 
In addition, we do not describe in detail the analysis of HXD-GSO data, 
since the HXD-GSO detected no significant signal from the source.

\begin{table*}[htdp]
\caption{Log of {\it Suzaku} Observations \label{tab:ls5039_obs}
}
\begin{center}
\begin{tabular}{cccc}
\hline
\hline 
Obs. ID & Coord. (J2000) & Exposure & Date \\ 
 & RA, DEC & XIS/HXD & \\
\hline
402015010 & $18^{\rm h}26^{\rm m}15^{\rm s}.1$, $-14^{\rm d}50^{\rm m}54^{\rm s}.2$ & 203~ks/181~ks & 09$-$15/09/2007\\
\hline
\end{tabular}
\end{center}

Note.---The exposure time is the net integration time after standard data screening for the XIS and HXD-PIN.
\end{table*}%

\begin{figure}[htdp]
\epsscale{1.0}
\plotone{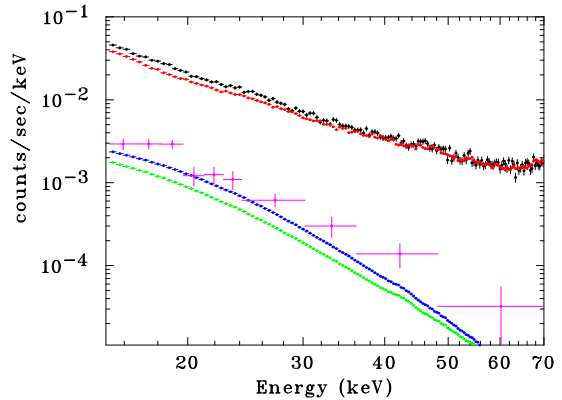}
\caption{The HXD-PIN spectrum. The black, red, blue, green, and magenta points show the count rates based on the raw data, 
from the modeling of NXB, the CXB component, the GRXE component, and the background-subtracted data, respectively. 
The background here includes contributions from the NXB, CXB and GRXE  (see \S2 for details).}
\label{fig:draw_XSPEC_PIN_compare}
\end{figure}

\section{Data Reduction}

We used data sets processed using the software of the {\it Suzaku} data processing pipeline (version 2.1.6.16). 
Reduction and analysis of the data were performed following the standard procedure using the HEADAS v6.4 software 
package, and spectral fitting was performed with XSPEC v.11.3.2. 

For the XIS data analysis, we accumulated cleaned events over good time intervals that were selected by removing 
spacecraft passages through the South Atlantic Anomaly (SAA). Further, we screened the data with the following criteria --- 
(1) cut-off rigidity is larger than 6~GV and (2) elevation angle from the Earth's rim is larger than 5$^{\circ}$. 
The source photons were accumulated from a circular region with a radius of $3^{\prime}$.
The background region was chosen in the same field of view with the same radius and an offset of 
$9^{\prime}$ from the source region. 
We have co-added the data from the two FI-CCDs (XIS0 and XIS3) to increase statistics.
The response (RMF) files and the auxiliary response (ARF) files used in this paper 
were produced using {\tt xisrmfgen} and {\tt xissimarfgen}, respectively.

For the HXD data,``uncleaned event files'' were screened with the standard event screening criteria: 
the cut-off rigidity is larger than 6~GV, the elapsed time after the passage of the SAA 
is more than 500~s and the time to the next SAA passage is more than 180~s, high voltages from all eight HV units are within 
the normal range and the elevation angle from the Earth's rim is more than $5^{\circ}$. We also discarded telemetry-saturated 
time intervals. 
In the spectral analysis in \S3, we used the response file for a point-like source at the HXD-nominal 
position ({\tt ae\_hxd\_pinhxnome4\_20070914.rsp}), which is released as a part of CALDB ({\it Suzaku} calibration data base). 

The HXD-PIN spectrum is dominated by the time-variable instrumental background
 ( non-X-ray background (NXB) ) induced by cosmic rays and trapped charged particles in the satellite orbit. 
To estimate the instrumental background component, we used the the time-dependent  NXB
event files released by the HXD instrument team, whose reproducibility is 
reported by \cite{fuka08}. 
In order to estimate the systematic uncertainty of the NXB model during our observation, 
we compare the NXB model spectrum during Earth occultation 
with the observed spectrum of the same time interval. 
The event selection criteria for this study are the same as those of the cleaned event 
except for the criterion on Earth elevation angle, which was chosen to 
to be less than $-5^\circ$. The estimated uncertainty obtained is $\sim 3$\%, 
which is consistent with the values reported by \cite{fuka08}. 

Another component of the HXD-PIN background is the cosmic X-ray background (CXB).
In our analysis, we assumed the CXB spectrum reported by \citep{gruber99}:
\begin{equation}
I(\epsilon) = 7.9 \;  \epsilon^{-1.29}_{\rm keV}  \exp{\left( -\frac{\epsilon_{{\rm keV}}}{\epsilon_{\rm p}} \right)}~{\rm ph}~{\rm s}^{-1}~{\rm keV}^{-1}~{\rm cm}^{-2}~{\rm str}^{-1}, 
\end{equation}
where $\epsilon_{\rm keV} = \epsilon/{\rm keV}$ and $\epsilon_{\rm p} = 41.1$. 
The CXB spectrum observed with HXD-PIN was simulated by using a PIN response file 
for isotropic diffuse emission ({\tt ae\_hxd\_pinflate4\_20070914.rsp}) and added to 
the NXB spectrum. 
Based on this approach, the contribution from the CXB flux is $\sim 5$\% of the NXB. 

Since {\LS} is located close to the Galactic plane, the contribution from the 
Galactic ridge X-ray emission (GRXE) must be examined, especially for 
HXD-PIN spectra. In order to model the shape of the GRXE,  data from 
{\it Suzaku} observations of the Galactic ridge region (ObsID: 500009010 and 500009020) 
were analyzed. 
The {\it Suzaku} spectrum from 3~keV  to 50~keV can be well fitted with the Raymond-Smith plasma 
with a temperature of $kT = 2.2 \pm 0.8$~keV and a power-law function with 
$\Gamma = 1.92^{+0.28}_{-0.4}$.  
Although we also tried a power law with exponential cutoff, following the results from the 
{\it INTEGRAL} IBIS \citep{klivonos07}, it turned out that the assumption on the spectral shape 
of the GRXE has negligible effect on the spectral parameters of {\LS}. 
The normalization of the GRXE spectrum component in the HXD-PIN spectrum of {\LS} 
is determined from the XIS spectrum of the {\LS} observation by excluding an encircled region with a radius of 
$4.5^{\prime}$ centered on the {\LS} location. 
The flux of the GRXE is estimated to be $\sim 40\%$ of the contribution from the CXB. 
In Figure~\ref{fig:draw_XSPEC_PIN_compare}, we show the time averaged HXD-PIN spectrum plotted 
together with models for the NXB, CXB and GRXE.

\section{Analysis and Results}

\subsection{Temporal analysis}

\begin{figure}
\epsscale{1.0}
%\plotone{draw_lc_top_mod2.eps}
\plotone{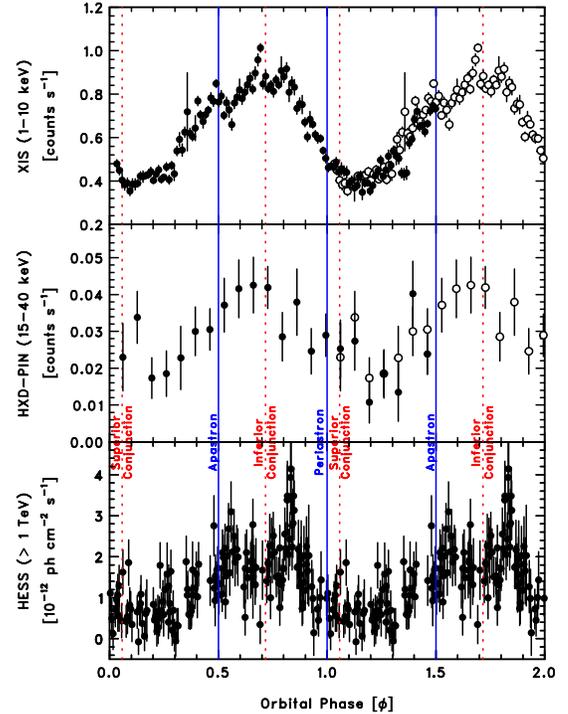}
\caption{Orbital light curves observed for \LS .
(Top) {\it Suzaku} XIS count rate 
in the 1--10 keV band continuously obtained for a 200~ks duration 
(filled circles), which covers one orbit and a half. 
Overlaid is the same light curve but shifted by one orbital period (open circles).
(Middle)  {\it Suzaku} PIN count rate 
in the 15--40 keV band. Background components consisting of NXB, CXB and GRXE are subtracted. 
(Bottom) the light curve of integral fluxes at energies E$>$ 1 TeV vs. orbital phase ($\phi$) obtained on run-by-run basis from the HESS data (2004 to 2005) 
reported by \cite{HESS_06}.
\label{fig:light-curve}}
\end{figure}

The light curve obtained from the XIS detector is shown in 
the top panel of Figure~\ref{fig:light-curve}.  
The continuous coverage in X-rays, longer than the orbital period of the {\LS} system, 
reveals a smooth variation of a factor 2 in the 1--10 keV count rate.
The light curve is drawn over two orbital periods. 
The orbital phase is calculated with the period of 
3.90603 days, and $\phi = 0$ 
with reference epoch $T_0$ (${\rm HJD}-2400000.5=51942.59$) 
taken from \cite{casares05}. 
The light curve from phase $\phi =1.0$ to 1.5, which was obtained in the last part of 
the observation, smoothly overlaps with the one obtained at the beginning of the
observation ($\phi = 0.0$--0.5). 

In the middle panel of Figure~\ref{fig:light-curve}, we present the light curve 
obtained with the HXD-PIN for the energy range 15--40~keV. 
Although the statistical errors are larger, the modulation behavior 
is similar to that of the XIS. The amplitude of the modulation is 
roughly the same between the XIS and HXD-PIN, indicating 
small changes of spectral shape depending on orbital phase. 
The spectral parameters obtained for each orbital phase 
are reported in the following section. 

The light curves obtained with {\it Suzaku} show that 
the X-ray flux  minimum appears around phase 0.0--0.3 and it reaches maximum around phase 0.5--0.8. 
 In order to quantify the amplitude of the flux variations, we fitted the XIS  light curve
with  a simple sinusoidal function.  Due to structures in the light curve,
 the fit converges with large chi-square ($\chi_{\nu}^2(\nu)$ = 4.92 (121)).  However,
 the general trend is well represented by a sinusoidal function:
 \begin{equation}
  I (\phi) = 0.24 \sin (\phi - 0.13) + 0.64 \; {\rm counts} \; {\rm s}^{\rm -1}
 \end{equation}
 where $I(\phi)$ is the count rate as a function of phase $\phi$.
 The ratios between the minimum and maximum count rates are 2.21 $^{+0.02}_{-0.03}$ counts s$^{-1}$ for XIS and 2.02 $^{+0.25}_{-0.19}$ counts s$^{-1}$ for HXD-PIN.
Structures of the X-ray and hard X-ray light curves are similar to that discovered in the phase diagram of integral fluxes at energies $>1$~TeV 
obtained on a run-by-run basis from HESS data (2004 to 2005) \citep{HESS_06}. The temporal X-ray behavior 
was already suggested by {\it RXTE} data, as well as by a compilation of all the previous X-ray data obtained with imaging 
instruments \citep{vbosch05, zabalza08}. 

In addition to the continuously changing component with respect
to the orbital phase, short timescale structures are found around $\phi=0.48$ and $\phi=0.7$. 
The unabsorbed flux changes about a 30\% in  $\Delta\phi \sim 0.05~(=4.7~{\rm hour})$. 
A significant dip can be seen around $\phi=1.35$ in the top panel of Figure \ref{fig:light-curve}. 
In comparison with the data at  around  $\phi=0.35$  obtained in the first half of the observation, 
the flux decreased $\sim$50\% only in this phase.  
The time duration of this dip corresponds to $\Delta\phi\sim0.03$. 
These structures may reflect features of the (possibly changing) environment of the X-ray emitting region, it is therefore 
of importance to test with further observations if these are persistent features.  

\subsection{Spectral Analysis}

Firstly we study time-resolved (phase-resolved) X-ray spectra.  
The data are divided into data segments with respect to the assigned phase, 
and model fitting is performed for XIS spectra 
for each segment with $\Delta \phi = 0.1$.

A single power-law function with photoelectric absorption, 
provides a good fit for all the segments. 
In order to study the possible changes of the 
amount of photoelectric absorption, we here fit the data
with $N_{\rm H}$ free. 
The best-fit parameters are presented in Table \ref{tab:ls5039_xis}. 
The derived values of the photon index and absorption column density are consistent with previous observations 
\citep{martocchia05, vbosch07}.  When we fix the $N_{\rm H}$ to the value obtained from the time
averaged spectrum,  resultant photon indices stay same within statistical error.

\begin{table*}[htdp]
\caption{Results of {\it Suzaku} XIS spectral fitting}
\label{tab:ls5039_xis}
\begin{center}
\begin{tabular}{ccccc}
\hline
\hline
Orbital Phase Intervals  & Photon index & $N_H~(wabs)$ &Flux [1-10~keV] & $\chi^2_{\nu}(\nu)$\\
& & $10^{21}~{\rm cm}^{-2}$ & $10^{-12}{\rm erg}~{\rm cm}^{-2}~{\rm s}^{-1}$ \\
\hline
INFC ($0.45<\phi\leq0.9$) & 1.48 $\pm$   0.02 &  7.82 $\pm$ 0.18 &   10.78 $\pm$ 0.05     & 0.97 (111) \\
SUPC($\phi\leq0.45$, ~$0.9<\phi$) &  1.55$\pm$ 0.02 & 7.77 $\pm$ 0.20 & 6.72 $\pm$  0.04 & 0.94 (111)\\
All Phases (time averaged) & 1.51$\pm$0.01 & 7.71$\pm$0.2 &  8.07 $\pm$ 0.03  & 0.81 (118)\\
\hline
0.0-0.1	& 1.57 $\pm$ 0.04	& 	6.86 $\pm$ 0.44	& 5.62 $\pm$ 0.03 & 0.99 (111)\\
0.1-0.2	& 1.61 $\pm$ 0.04	& 	7.67 $\pm$ 0.46  	& 5.18 $\pm$ 0.03  & 1.04 (111)\\
0.2-0.3	& 1.51 $\pm$ 0.03	& 	7.23 $\pm$ 0.37 	& 5.67 $\pm$  0.02 & 1.13 (107)\\
0.3-0.4	& 1.49 $\pm$ 0.03	& 	7.42 $\pm$  0.43	& 7.34 $\pm$ 0.03   & 1.01 (111)\\
0.4-0.5	& 1.45 $\pm$ 0.02	& 	7.63 $\pm$ 0.26 	& 9.73 $\pm$  0.01  & 0.96 (111)\\
0.5-0.6	& 1.46 $\pm$ 0.03	& 	7.72 $\pm$  0.45	& 9.95 $\pm$ 0.02    & 1.15 (105)\\
0.6-0.7	& 1.46 $\pm$ 0.03	& 	7.82 $\pm$  0.37	& 12.05 $\pm$ 0.02  & 0.77 (118)\\
0.7-0.8	& 1.51 $\pm$ 0.02 	& 	7.61 $\pm$ 0.33	& 11.27 $\pm$ 0.02  & 1.10 (108)\\
0.8-0.9	& 1.52 $\pm$ 0.04	& 	7.63 $\pm$  0.47	& 10.29 $\pm$ 0.03 & 0.87 (111)\\
0.9-1.0	& 1.59 $\pm$ 0.03	& 	8.35 $\pm$ 0.41	& 7.84  $\pm$ 0.02  & 1.06 (111)\\
\hline
\end{tabular}
\end{center}
Note.---Fitting {\it Suzaku} XIS0 +XIS3 spectrum of LS5039 by a power-law with photoelectric 
absorption in 0.6--10~keV. Photon index $\Gamma$, 
absorbing column density  $N_{\rm {H}}$, and the $1-10$ keV flux $F_{\rm{1-10}}$ (corrected for absorption) are shown with 68.3 \% (1~$\sigma$)
error. The orbital phase $\phi$ is calculated from the ephemeris of \cite{casares05}.
\end{table*}

\begin{figure}
\epsscale{1.0}
%\plotone{lc_LS5039_XIS_081208.eps}
\plotone{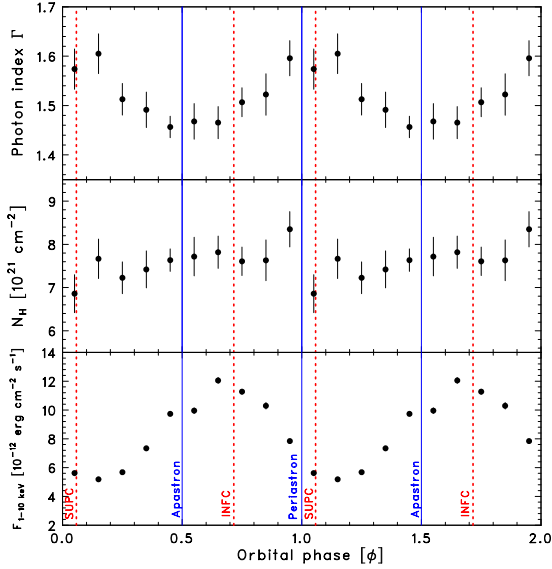}
\caption{Orbital light curves of the photon indices, 
$N_H$, and unabsorbed fluxes in the energy range of 1-10~keV obtained by fitting with an absorbed power-law model. 
The phase interval width is $\Delta\phi=0.1$. Two full phase periods are shown for clarity. The blue solid lines show 
periastron and apastron phase and red dashed lines show SUPC and INFC of the compact object.}
\label{fig:lc_LS5039_XIS_080923}
\end{figure}

The photon index ($\Gamma$) values are plotted as a function of orbital phase 
in the  top panel of Figure~\ref{fig:lc_LS5039_XIS_080923}. 
The spectral shape varied such that the spectrum is steep around SUPC
($\Gamma \simeq  1.61$) and becomes hard ($\Gamma \simeq 1.45$)
around apastron. 
The modulation behavior of $\Gamma$ is somewhat different from 
that observed using HESS in the VHE range. 
The amplitude of the variation is $\pm 0.1$, which is much smaller 
than the change of $\pm$0.6 in the VHE region \citep{HESS_06}.  
The 1--10 keV flux changes from $(5.18 \pm 0.03)\times 10^{-12}~{\rm erg}~{\rm cm}^{-2}~{\rm s}^{-1}$
($\phi= 0.1\mbox{--}0.2$) 
to $(12.05 \pm 0.02) \times 10^{-12}\ \rm erg\  cm^{-2}\ s^{-1}$. ($\phi= 0.6\mbox{--}0.7$).

In all the data segments, the source is significantly detected with the HXD-PIN, 
indicating that hard X-ray emission extends at least up to 70 keV.   
Note also that although the XIS and HXD-PIN spectra do not overlap,  
they seem to be smoothly connected in the gap between 10 and 15 keV.

\begin{figure}[tbhp]
\epsscale{1.0}
\begin{center}
\plotone{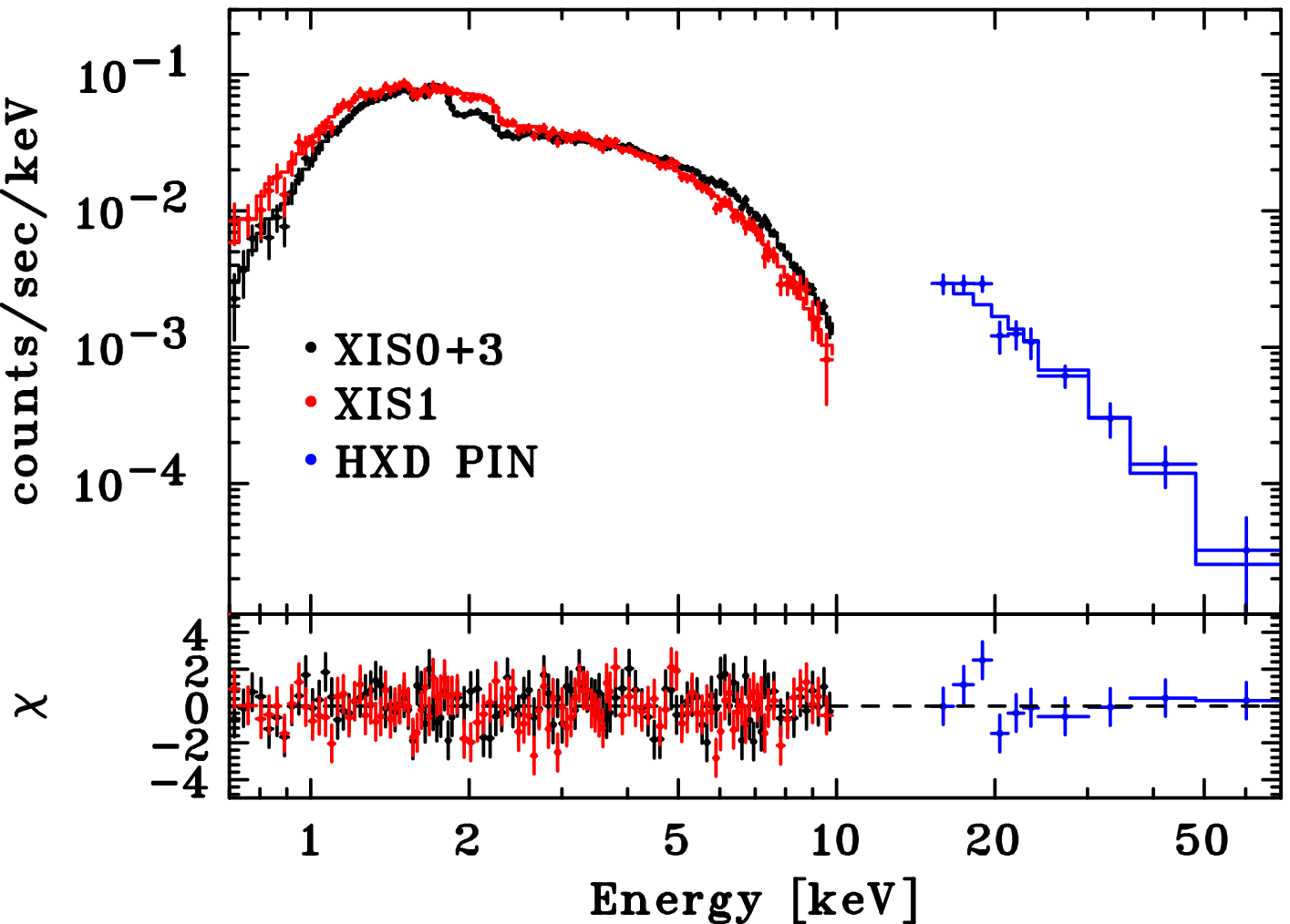}
\plotone{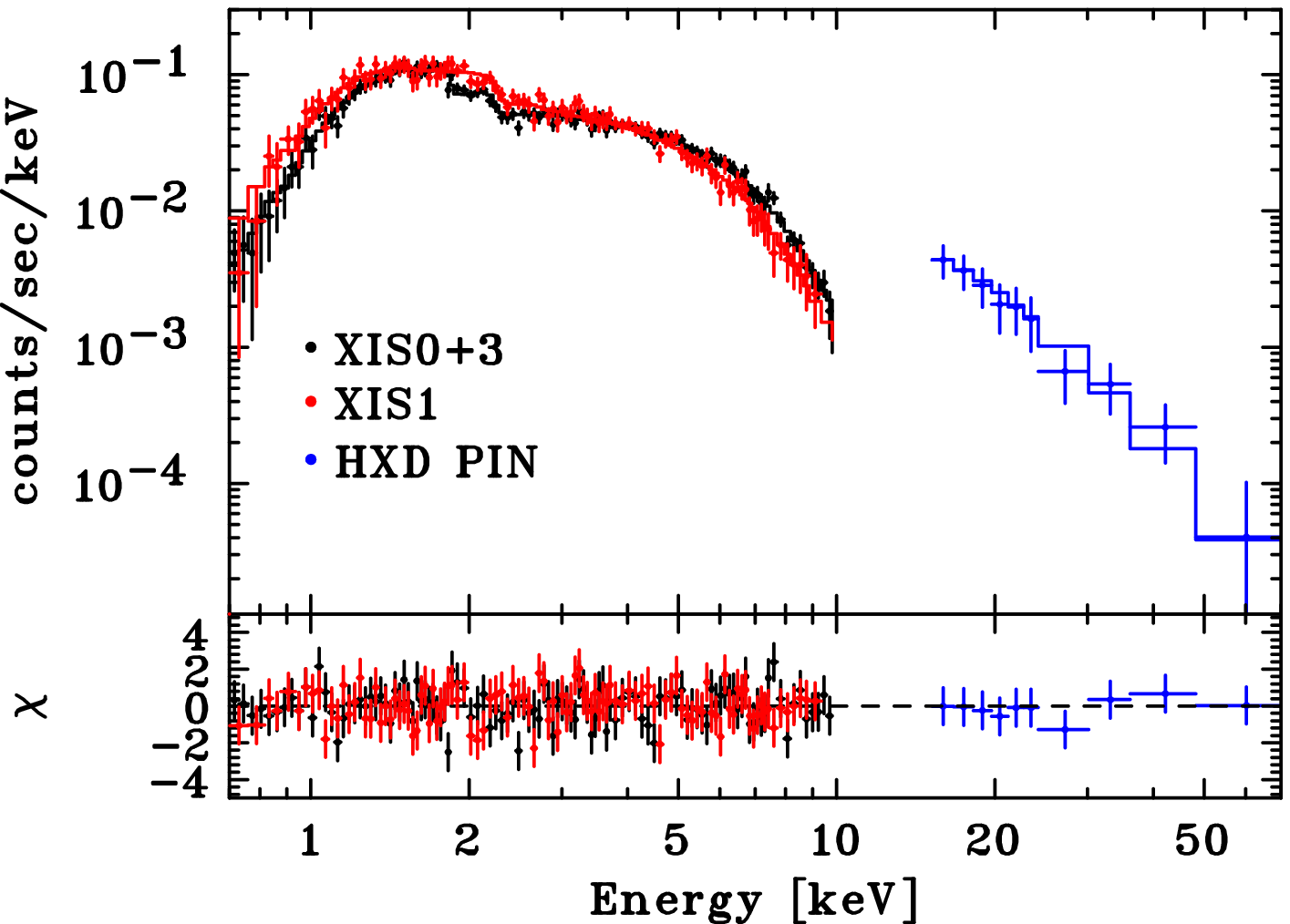}
\plotone{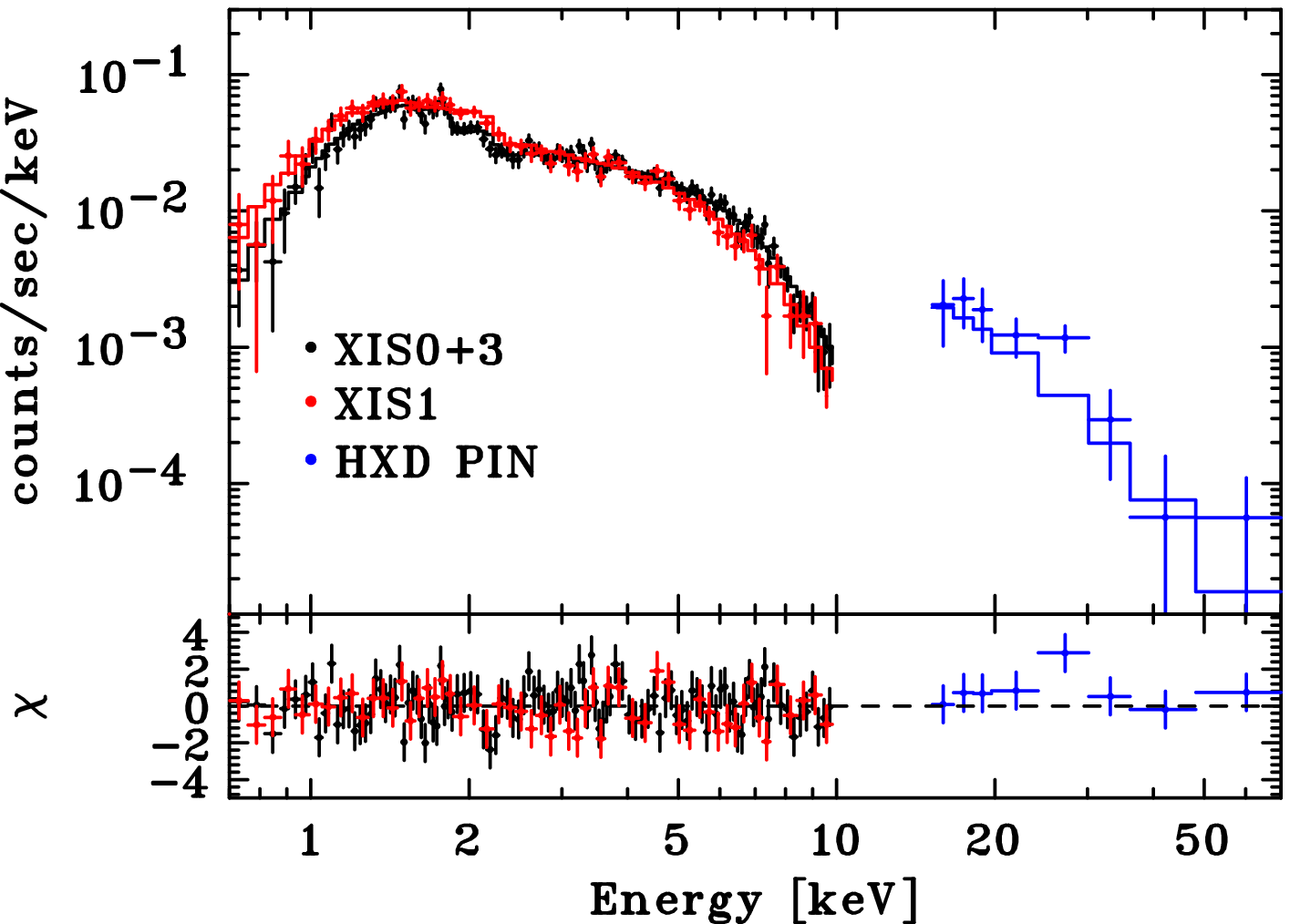}
\end{center}
\caption{Top: The time-averaged $Suzaku$ (XIS+HXD) spectrum  (Top) and spectra obtained in INFC (Middle) and SUPC (Bottom).
\label{fig:draw_XSPEC_INFC_081108_mod}}
\end{figure}

To study the shape of the spectrum above 10 keV, the XIS spectra and the PIN spectrum in the range 15--70 keV
after subtraction of background (NXB $+$ CXB $+$ GRXE) are jointly fitted (Figure \ref{fig:draw_XSPEC_INFC_081108_mod} Top).
The  time-averaged spectra  are well represented by an absorbed power-law model with $\Gamma$= 1.51 $\pm$0.02 with reduced $\chi^2_{\nu}=0.99$ 
(235 degrees of freedom). We find no cutoff structure in the energy range of the HXD-PIN. 
The spectra  within the phase intervals [$0.616<\phi<0.816$] and [$\phi<0.158$ \& $0.958<\phi$], 
which correspond to the INFC and SUPC, respectively, are also shown 
in Figure \ref{fig:draw_XSPEC_INFC_081108_mod}. The best-fit parameters are presented in Table \ref{tab:ls5039_xis_pin}. 

\begin{table*}[htdp]
\caption{Results of {\LS} XIS+HXD-PIN spectral fitting
\label{tab:ls5039_xis_pin}}
\begin{center}
\begin{tabular}{cccccc}
\hline
\hline 
Orbital Phase Intervals &  & Photon index & $N_H~(wabs)$ & $\chi_{\mu}^2(\nu)$ \\
& & &  $\times10^{21}~{\rm cm}^2$ & & \\
\hline
All Phases (time averaged) & & 1.51$\pm$ 0.02 & 7.7 $\pm$ 0.2 & 0.95   (175 d.o.f.) \\
INFC ($0.616<\phi<0.816$) && 1.49 $\pm$ 0.04 & 7.9 $\pm$ 0.4 & 0.87    \\
SUPC ($\phi<0.158, 0.958<\phi<1.158$)& & 1.55 $\pm$ 0.05 & 6.9 $\pm$ 0.5 & 1.06    \\
\hline
\end{tabular}
\end{center}

Note.---Power-law fitting to the XIS/HXD-PIN spectrum. The errors of spectral parameters 
are at the 90\% confidence level. 
\end{table*}%

 Althought earlier observations by {\it RXTE} suggested the presence of an 
iron emission line at 6.7 keV (Ribo et al. 1999), later observations by 
{\it Chandra} and {\it XMM} could not find evidence of it (e.g. 
\citealt{vbosch05}).  A careful study of new and longer {\it RXTE} 
observations, using slew data to account for background emission, revealed 
that the earlierly reported 6.7 keV emission line is likely a background 
feature  \citep{vbosch05} . The {\it Suzaku} data confirm this result.
In an attempt to find the possible signature of iron emission lines from {\LS}, 
we analyzed the phase-averaged spectrum.
The upper limits on iron line structures are determined by fitting a Gaussian 
at various energies and line widths at which Fe emission might be expected. 
The power-law continuum model parameters are fixed with the best fit
values and a Gaussian component is added to the power-law function. 
The central energy of the Gaussian line is swept from 6.0~keV to 7.1~keV in steps of 0.1~keV. Line widths are changed
from 0.01~keV to 0.09~keV in steps of 0.01~keV, together with lines with larger widths of 0.15 and 0.20~keV. 
An equivalent width is determined at each grid point. The resulting upper-limit on the equivalent width is 40~eV
 with 90\% confidence level.

\section{Discussion}

The X-ray emission observed with {\it Suzaku} is characterized by 
(1) a hard power law with $\Gamma \simeq 1.5$ extending from soft X-rays to $\sim 70$ keV, 
(2) clear orbital modulation in flux and photon index, 
(3) a moderate X-ray luminosity of 
$L_X \sim 10^{33} (\frac{D}{1 \rm kpc})^2\ \rm erg\ s^{-1}$,
(4) a small and constant absorbing column density, and
(5) a lack of detectable emission lines.

Although variable X-ray emission has been found from more than two hundred  galactic binary systems, the {\it Suzaku} data
hardly can be explained within the ``standard accretion''  scenario where  X-rays are produced by a hot thermal (comptonized)  accretion
plasma around the compact object.  The lack of X-ray emission lines (at the level of sensitivity of {\em Suzaku})  
as well as  the hard $E^{-1.5}$ type energy spectrum of  the X-ray continuum, extending from soft X-rays up to 70~keV,
favors  a non-thermal origin of the X-rays. 
This conclusion  is supported by the general similarities between the properties of the observed X-rays and TeV gamma-rays. 
Namely,  both radiation components  require a rather hard energy distribution of parent electrons with a power-law  index of
$\approx 2$.  This directly follows from the photon index of the synchrotron radiation  $\Gamma \approx 1.5$, and agrees
quite well with the currently most favored interpretation of the TeV gamma-rays, in which they would be produced by IC scattering
off  the anisotropic photon field of the massive companion star.  

\begin{figure}[htbp]
\begin{center}
\epsscale{1.0}
\plotone{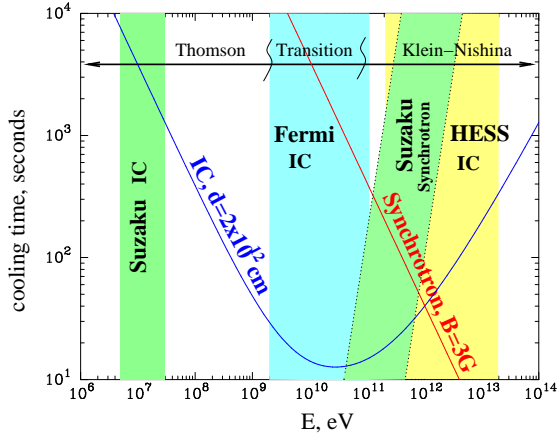}
\caption{
The radiative cooling times as a function of electron energy. The blue line
corresponds to IC losses at the distance $d=2\times10^{12}$~cm from the optical
star (the star luminously and temperature were assumed to be
$L_*=7 \times 10^{38}$~erg/s and $T_*=3.8\times10^{4}$~K, respectively). The
red line corresponds to the synchrotron losses for B-field $B=3$~G. The
filled regions reproduce the electron energy intervals relevant to the HESS
(yellow),  {\it Fermi} (light blue) and {\it Suzaku}  (green) energy domains.
Formally, 
two radiation channels, synchrotron radiation of 
very high energy electrons (light green) and 
IC scattering  of  low energy electrons in the Thompson regime (dark green)   
can produce X-ray photons in the {\it Suzaku} energy domain.}
\label{fig:diagram}
\end{center}
\end{figure}

Assuming that the TeV gamma-ray production region is located at a
distance from the companion star of $d \sim 2\times 10^{12}$~cm (i.e. the binary system size),  and taking into account that
gamma-rays are produced in the deep Klein-Nishina (KN) regime with significantly suppressed cross-section,  for the well
known  luminosity of the  optical star $L \simeq  7 \times 10^{38}\,{\rm erg}\,{\rm s}^{-1}$, one can estimate quite 
robustly the strength of the magnetic field in the emission region. The numerical  calculations show that the field  should 
be around a few Gauss (see e.g. Fig. 5). For such a magnetic field strength, the  energy intervals of  electrons responsible
for the two emission components overlap substantially, as shown in Figure~\ref{fig:diagram}. Therefore, we are most likely
dealing with the same  population of parent electrons, which should be located  at large distances from the compact object,
in the system periphery,  to prevent the severe absorption of the TeV radiation and the subsequent intense emission from the
pair-created secondaries \citep{Mitya_LS, vboshch08}.

It should be noted that the observed X-ray emission is very difficult to explain as synchrotron emission produced by secondary
(pair-produced) electron and positrons. Since the pair production cross-section has strong energy dependence with a distinct
maximum, for the  target photons of typical energy of  $\sim 10$~eV, the major fraction of the absorbed energy will be 
released in the form of $\sim 100$~GeV electrons. Thus secondary pair synchrotron emission must show a spectral break in the
{\it Suzaku} energy band unless one assumes unreasonably high magnetic fields, $B \geq 1$~kG, in the surroundings of the gamma-ray
emission region \citep{Mitya_LS,vboshch08}. 

Figure~\ref{fig:diagram} shows the synchrotron and IC cooling times of electrons, as a function of electron energy, 
calculated for the  stellar photon  density  at  $d = 2\times 10^{12}\ \rm cm$ and for a magnetic field $B=3$~G. It can be seen
that synchrotron losses dominate over IC losses at $E_{\rm e} \geq 1$~TeV.  Note that  the  TeV gamma-ray production takes
place in the deep KN regime. This implies that the cooling time, $t_{\rm cool} = \gamma/\dot{\gamma}$, of electrons
generating GeV gamma-rays via IC scattering (Thomson regime) is shorter than the cooling time of TeV electrons responsible
for producing very high-energy gamma-rays (KN). The same applies for synchrotron cooling time of multi-TeV electrons  that
produce low-energy (MeV) gamma-rays by synchrotron radiation.  One should therefore expect significantly higher  MeV
(synchrotron) and GeV (IC) fluxes than at keV and TeV energies, provided that the acceleration spectrum of electrons 
extends from low energies to very high energies.  However,  in the case of  existence of low-energy and very-high-energy
cutoffs in the acceleration spectrum, the gamma-ray fluxes $>10$~MeV and at GeV energies would be significantly suppressed. 

To better understand the energy ranges of the electrons responsible 
for X-ray and gamma-ray production, we show in Figure~\ref{fig:diagram} the energy zones of
electrons relevant to the {\it Suzaku}, {\it Fermi}, 
and HESS radiation domains. Note that the reconstruction 
of the average energy of electrons responsible for the IC gamma-rays 
depends only on the well known  temperature of the companion star $T=3.8\times 10^4$~K.
The light green zone in Figure~\ref{fig:diagram}  marked as ``{\it Suzaku} synchrotron'', 
corresponds to electrons responsible  for the synchrotron photons produced in the
energy interval
$1~{\rm keV}\le \epsilon_{\rm syn}\le40~{\rm keV}$. 
For a reasonable range of magnetic field values,
the energy interval of electrons relevant for {\it Suzaku} 
data overlap on one hand with the HESS energy interval, and can overlap with the {\it
Fermi} one. This 
should allow us, in the case of detection of MeV/GeV gamma-rays by {\it Fermi}, to
considerably reduce the parameter space, 
in particular, to better localize the X- and 
gamma-ray production regions from electromagnetic cascade constraints, and derive the broadband energy
spectrum of electrons and the strength of the magnetic field, both as a function of the orbital phase.  

Formally, when X-rays and TeV gamma-rays are produced by 
the same population of  very-high-energy electrons,  one should expect a general 
correlation between the light curves obtained  by {\it Suzaku} and HESS.  
In this regard, the  similarity between the {\it Suzaku} and HESS light curves  
seems to be natural. However, such an interpretation is not
straightforward in 
the sense that two major mechanisms that might cause modulation of the TeV
gamma-ray signal are  
related to interactions of electrons and gamma-rays with the photons of the companion
star, i.e. 
anisotropic IC scattering and photon-photon pair production \citep{Mitya_LS, Dubus2008}, 
and thus cannot contribute to the  
X-ray modulation. The X-ray modulation requires periodic changes of the 
strength of the ambient magnetic field or the number of relativistic electrons. Note,
however, 
that the change of magnetic  field would not have a strong impact as long as the
radiation 
proceeds in the saturation regime and
synchrotron losses dominate in the relevant energy interval. One would also expect
modulation of the synchrotron X-ray flux if the energy losses of electrons are 
dominated by IC scattering, although in such a case we should observe
significantly lower X-ray fluxes.

A more natural reason for the modulation of the synchrotron fluxes would come from dominantly adiabatic losses. The adiabatic cooling
of electrons  in binary systems can be realized through complex (magneto)hydrodynamical  processes, e.g. due to interactions 
between a black hole jet or a pulsar wind with the dense stellar wind of a massive companion star (see e.g.
\citealt{Bogovalov08}, \citealt{perucho08}). The orbital motion could naturally produce the modulation of  adiabatic cooling of
electrons around the orbit (see e.g. \citealt{khangulyan_dublin}). Note that because of the relatively small  variation of the
X-ray flux  over the orbit, a factor of only two, the requirements for this scenario are quite modest.
We note that dominant adiabatic losses have been invoked by
\cite{Khangulyan2007} to explain the variations of the X-ray and TeV gamma-ray  fluxes from the binary pulsar {\PSR}. 

The detected power-law spectrum of X-rays with photon index around $\Gamma =1.5$ implies  that the established energy
spectrum of electrons is also a power-law with index $\alpha_{\rm e}\simeq 2$. 
This agrees well with the hypothesis of
dominance of adiabatic losses, because the  adiabatic losses do not change the initial spectrum of electrons. Thus the
required power-law index $\alpha_{\rm e}\simeq 2$ implies  a reasonable acceleration spectrum $Q(E_{\rm e})\propto E_{\rm e}^{-2}$.
Otherwise, in an environment dominated by synchrotron losses, the acceleration spectrum should be very hard, with a power-law index $\leq 1$, or should have an unreasonably large low-energy cutoff at $E \geq 1$~TeV to explain the
observed X-ray spectra.   

Obviously, adiabatic losses modulate the IC gamma-ray flux in a similar manner. However, unlike X-rays, the TeV gamma-rays suffer significant distortion due to photon-photon absorption (see
e.g. \citealt{boettcher08}) and anisotropic IC scattering with its strong hardening of the gamma-ray spectrum \citep{mitya_felix05}. All this leads to additional orbital modulation of the gamma-ray signal, and it is likely that these
two additional processes are responsible for the strong change of gamma-ray flux, much more pronounced than that seen
in X-rays (see Fig.~\ref{fig:light-curve}). The {\it Suzaku} data  presented in  this paper implies a key additional
assumption, namely that the accelerated electrons must loose their energy adiabatically before they cool
radiatively.

In order to demonstrate that the suggested scenario can satisfactorily explain the combined  {\it Suzaku} X-ray and 
HESS  gamma-ray data,  we performed  calculations of  the broad-band  spectral energy distributions (SEDs) of the synchrotron
and IC emission, assuming a simple model in which the same population of electrons is responsible for both X-rays and TeV
gamma-rays. We also assumed that the emission region has homogeneous physical conditions. This is a reasonable assumption 
given that we deal with  very  short  cooling timescales  ($\ll  100$~s), thus electrons cannot travel significant distances
while emitting. 

\begin{figure}[htbp]
\begin{center}
\epsscale{1.0}
% \plotone{ligt_curves.eps}
\plotone{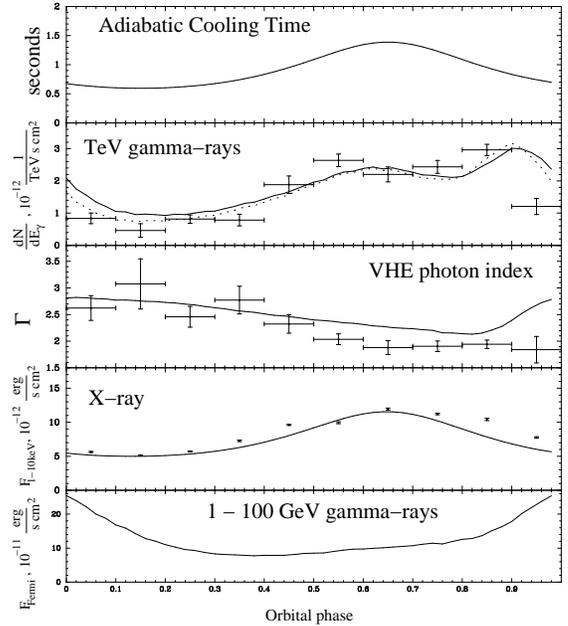}
\caption{Lightcurves.  (a) the phase dependence of adiabatic losses derived from the {\it Suzaku} data;
(b) solid line: calculated 1~TeV gamma-ray fluxes together with HESS data points, dotted line: the normalizations at 1~TeV of  power-law fits for the calculated spectra; (c)  calculated photon indices  around 1~TeV  together with power-law indices reported by HESS; 
(d) calculated 1--10 keV   X-ray fluxes  with the reported {\it Suzaku} data; 
and 
(e) predicted 1-100 GeV integrated IC flux (note that the peak 
in the spectrum is at 10 GeV; see Fig. 7).  The calculations were
performed for the following parameters: inclination of the orbit
$i=30$~deg; elevation over the orbital plane $Z_0=1.8 \times 10^{12}$~cm;
magnetic field strength $B=2.5$~G. A power-law injection spectrum of electrons 
$Q(E) \propto E_{\rm e}^{-2}$ ($10{\rm GeV}<E_{\rm e}<100{\rm TeV}$) was assumed.
The total power injected in relativistic electrons was fixed at the level of $10^{37}$~erg/s.
The cooled electron spectrum is formed due to radiative (IC and synchrotron)
and non-radiative (adiabatic) losses, derived from the  {\it Suzaku} data.
\label{fig:phase} }
\end{center}
\end{figure}

In the regime dominated by  adiabatic energy losses,  the  synchrotron X-ray flux is proportional to $t_{\rm ad}$. The
X-ray modulation seen by {\it Suzaku} is then described by  the modulation of the adiabatic loss rate. 
In Fig.~\ref{fig:phase}, we show $t_{\rm ad}(\phi)$ that is inferred from
the X-ray data. 
The required adiabatic cooling timescales are $\sim 1$~s. 
Any consistent calculation of the
adiabatic cooling requires the solution of the corresponding hydrodynamical problem, and one needs to know in detail
the nature of the source. At the present stage, we consider the 
simple example of adiabatic cooling in a relativistically expanding source. In such a case, the adiabatic loss rate can be written as: $\dot{\gamma}_{\rm
ad}(\phi, \gamma) = \gamma / t_{\rm ad}$ with  $ t_{\rm ad} \sim R/c  \simeq 3 R_{11} $~sec,  where  $R_{11} \equiv
R/(10^{11}\ \rm cm)$ is the characteristic size of the source. The required variation of the adiabatic cooling is thus
reduced to the modulation of the size of the radiation region ($R_{11} \sim 0.3\mbox{--}1$). The size in turn depends on
the external pressure exerted by, e.g., the stellar wind from the massive star. The expected weaker external pressure around
apastron implicitly assumed in our model would be broadly consistent with the radial dependence of the wind pressure. 

\begin{figure}[htbp]
\epsscale{1.0}
\begin{center}
%\plotone{sed_av.eps}
\plotone{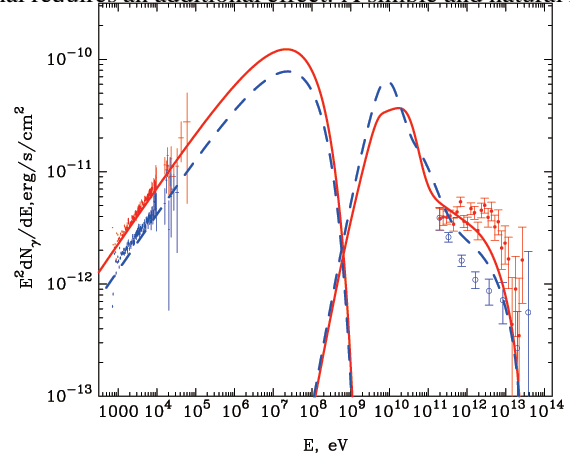}
\caption{
Model calculations of the non-thermal radiation spectra of  LS~5039, averaged
over SUPC (low state) and INFC (high state) orbital phase intervals 
for $\phi \leq 0.45$, $\phi \geq 0.9$ (SUPC, Blue); and
for $\phi=0.45$--$0.9$ (INFC,Red).
The flux points of 
$Suzaku$ X-ray spectra (1--40~keV band) are reconstructed with the best-fit function with correction
for the interstellar absorption.  The HESS gamma-ray spectra are taken from
Aharonian et.(2006). 
The calculations were performed for the same 
model parameters as in Fig. \ref{fig:phase}.
\label{fig:SED}} 
\end{center}
\end{figure}

In Fig.~\ref{fig:SED} we show the SEDs
averaged over the INFC ($\phi=0.45$--$0.9$) and SUPC
($\phi \leq 0.45$, $\phi \geq 0.9$) phase  intervals. 
The corresponding gamma-ray data
have been previously reported by HESS \citep{HESS_06}.
Since both the absolute flux and the energy spectrum of
TeV gamma-rays vary rapidly  with phase,  in order to compare the
theoretical predictions with  observations, we should use 
smaller phase bins, ideally
speaking with the time intervals  $\Delta t \leq 100$~s corresponding to
the characteristic cooling timescales of electrons. 
Because of the lack of the relevant gamma-ray data available to us,
we here 
 use the X-ray and gamma-ray data integrated over $\Delta \phi=0.45$.
While  this compromise does not allow us to
perform  quantitative studies, it can be used  to make a
qualitative comparison of the model calculations with observational data.

The theoretical calculations of the SEDs 
are in a reasonable agreement with the observed spectra, 
though they do not perfectly match the gamma-ray fluxes. 
One can improve the fits by
introducing slight phase-dependent changes in the spectra of accelerated
electrons, but it is beyond the scope of this paper given the 
caveat mentioned above. 
We should also note that the calculations of low energy gamma-rays 
(in the {\it Fermi} domain) 
are performed assuming that the injection spectrum of electrons continues down to $10$~GeV. If this is not the case, 
the gamma-ray fluxes in the {\it Fermi} energy domain
may be significantly suppressed. On the other hand, the detection of
gamma-rays by {\it  Fermi} would allow us to recover the spectrum of electrons in
a very broad energy interval, and thus distinguish between different
acceleration models. Another important feature in this scenario is that a hard 
synchrotron spectrum extending up to a few MeV, 
is required by the robust detection of $\sim10$~TeV gamma-rays from the system. 
A future detection of the emission at MeV energies 
may bring important information on the presence of highest energy particles in the system. 

The  reproduction, at least qualitatively, of the observed spectral and temporal features of the nonthermal radiation with
the simple toy model supports the production of X-ray and TeV gamma-rays by the same population of parent
particles and allows us to derive several principal conclusions. 
The electron energy distribution  should be a power-law with an almost constant index of $\alpha_{\rm e} = 2 \Gamma_{\rm X} -1
\simeq 2$ to explain the X-ray spectra. Note that in an \textit{isotropic} photon gas when the Compton scattering takes
place in the deep KN regime  such an electron distribution results in a quite steep TeV spectrum with photon index 
$\Gamma_\gamma \simeq  1+ \alpha_{\rm e} \simeq 3$. This does not agree with HESS observations. However, the anisotropic IC
provides a remarkable hardening of the gamma-ray spectrum \citep{mitya_felix05}, in particular, $\Gamma_\gamma \sim 2$ would
be expected for the INFC, and it has indeed been observed using HESS \citep{HESS_06}. We also note that, to explain the VHE
spectrum at SUPC, we have to assume that the emission region is located at a distance $\approx 2\times 10^{12}$~cm from the
compact object, in the direction perpendicular to the orbital plane. In the standard pulsar scenario, the production region
cannot be located far away from the compact object, and even invoking electromagnetic cascading (see e.g.  Fig. 16 in
\citealt{diego08}), one cannot reproduce the reported fluxes around  orbital phase 0.0
\citep{Mitya_LS, vboshch08}. 
We have found that the magnetic field strength cannot deviate much from a few G. We can also derive a constraint on the size of
the emission region,  imposing a maximum expansion speed of $\sim$$c$ (the speed of light), and the Hillas criterion
\citep{Hillas}, in which the minimum size of a source capable of accelerating particles to a given energy $E_{\rm e}$ is
$R=R_{\rm L}$ (where $R_{\rm L}=E_{\rm eV}/300 B_{\rm G}$~cm),  the Larmor radius. This estimate yields a size of
$10^{10}$--$10^{11}$~cm, which agrees quite well with the estimate based on the required timescale of adiabatic cooling. Note
that  the  requirement of fast adiabatic losses imposes a strong constraint on the acceleration rate of electrons. Indeed,
the acceleration timescale  can be expressed as: $t_{\rm acc} = \eta  R_{\rm L}/{c} \sim 0.1  \eta  ({E_{\rm e}}/{\rm 1
TeV})  ({B}/{\rm 1 \ G})^{-1} \ \rm s$,  where $\eta \geq 1$ parametrizes the acceleration efficiency. In extreme
accelerators with the maximum possible rate allowed by classical electrodynamics $\eta =1$. The HESS spectrum provides
evidence of electron acceleration well above 10~TeV. Therefore, $t_{\rm acc} < t_{\rm ad} \sim 1$~s  is required  at $E_{\rm
e} = 10$~TeV, which translates into $\eta <  3$  for  $B = 3$~G.  Thus we arrive at the conclusion that an extremely
efficient acceleration with $\eta < 3$  should operate in a compact region of $R \sim 10^{11}\ \rm cm$.

Finally, we would like to emphasize that in the scenario described here different radiation energy intervals are
characterized by fundamentally different light curves. While the synchrotron X-ray modulation is caused by  adiabatic losses,
the light curve in gamma-rays depends critically, in addition, on effects related to interactions with the
optical photons of the companion star. Two of these effects, photon-photon pair production and  anisotropic Compton
scattering are equally important for the formation of the light curve of TeV gamma-rays. On the other hand, only  the effect of
anisotropic Compton scattering has an impact on the formation of the light curve of GeV gamma-rays.  The difference of  light
curves in the X-ray and GeV and TeV gamma-ray intervals in this scenario is shown  in Figure \ref{fig:phase}.

\section{Summary}

The {\it Suzaku} X-ray satellite has observed {\LS} for the first time with imaging capabilities over one and a half orbits. 
 The {\it Suzaku} data show strong modulation of the
X-ray emission at the orbital period of the system and the X-ray spectral data are described by a hard power-law up to 70 keV.
We found the close correlation of the X-ray and
TeV  gamma-ray light curves, which  can be interpreted as evidence of production of these two radiation components  by the same 
electron population via synchrotron radiation and IC scattering, respectively. 
Whereas
there are at least two reasons for the formation of a periodic TeV gamma-ray light curve, both related to the interaction
with photons from the companion star (photon-photon absorption of VHE gamma-rays and IC  scattering in an anisotropic
photon field), the modulated X-ray signal requires an additional  effect.  A simple and natural reason for the modulation in
X-rays seems to be adiabatic  losses which should  dominate over the radiative (synchrotron and IC) losses of electrons.  We
demonstrate  that this assumption allows us to explain, at least qualitatively, the 
spectral and temporal characteristics of the 
combined {\it Suzaku} and HESS data. In particular, the introduction of adiabatic losses
not only provides a natural explanation  for the rather stable photon index of the X-ray spectrum $\Gamma_X \simeq
1.5$, but also allows one to approximately reproduce the TeV gamma-ray spectra. 

The gamma-ray data require a location of the production region at the periphery of the binary system at $d \sim 10^{12}$~cm.
This constraint allows a quite robust estimate of the magnetic field of a few Gauss to be derived directly from the X/TeV flux
ratio, and an adiabatic loss time of a few seconds to provide the dominance of adiabatic losses. In the case of a
relativistically expanding source, the size of the production region should not exceed $10^{11}$~cm. The adiabatic
cooling  cannot be shorter than several seconds, and correspondingly, the size of the production region cannot be much
smaller than $10^{11}$~cm, since otherwise the electrons could not be accelerated up to energies beyond 10~TeV, even 
assuming an extreme acceleration rate close to the fundamental limit determined by quantum electrodynamics.
 There is little doubt that  future simultaneous observations  of
{\LS} with the {\it Suzaku}, {\it Fermi}, and HESS telescopes will provide key information for understanding the nature of this mysterious non-thermal source.

\acknowledgments
T. Kishishita and T. Tanaka are supported by research fellowships of the 
Japan Society for the Promotion of Science for Young Scientists. 
 The authors acknowledge support by the Spanish DGI of MEC under grant 
AYA2007-6803407171-C03-01, as well as partial support by the European 
Regional Development Fund (ERDF/FEDER). V.B-R. acknowledges support by the 
Spanish DGI of MEC under grant AYA2007-6803407171-C03-01, as well as 
partial support by the European Regional Development Fund (ERDF/FEDER).


\begin{thebibliography}{}

\bibitem[Acciari et al.(2008)]{VERITAS_LSI} Acciari, V.~A., et al.\  2008, \apj, 679, 1427 

\bibitem[Aharonian et al.(2005a)]{HESS_5039} Aharonian, F., et al.\ 2005, Science, 309, 746 

\bibitem[Aharonian et al.(2005b)]{HESS_1259} Aharonian, F. A., et al.\ 2005, \aap, 442, 1

\bibitem[Aharonian et al.(2006)]{HESS_06} Aharonian, F., et al.\ 2006, \aap, 460, 743 

\bibitem[Albert et al.(2006)]{MAGIC_LSI} Albert, J., et al.\ 2006, Science, 312, 1771 

\bibitem[Albert et al.(2007)]{MAGIC_CygX1} Albert, J., et al.\ 2007, \apj, 665, L51 

\bibitem[Bogovalov et al.(2008)]{Bogovalov08} Bogovalov, S.~V., 
Khangulyan, D.~V., Koldoba, A.~V., Ustyugova, G.~V., \& Aharonian, F.~A.\ 
2008, \mnras, 387, 63 

\bibitem[Bolton(1972)]{Bolton72} Bolton, C.~T.\ 1972, \nat, 235, 271 

\bibitem[Bosch-Ramon et al.(2005)]{vbosch05} Bosch-Ramon, V., Paredes, J. M., Rib\'o, M. et al. 2005, 
\apj, 628, 388 

\bibitem[Bosch-Ramon et al.(2007)]{vbosch07}  Bosch-Ramon,~V., et al. 2007, \aap, 473, 545

\bibitem[Bosch-Ramon et al.(2008a)]{vboshch08} Bosch-Ramon, V., Khangulyan, D, \& Aharonian, F. A. 2008a, 
\aap, 482, 397

\bibitem[B\"ottcher(2007)]{boettcher08} B\"ottcher, M. 2007, Astr. Phys., 27, 278

\bibitem[Casares et al.(2005)]{casares05} Casares,~J., et al.\ 2005, \mnras, 364, 899

\bibitem[Dubus(2006)]{Dubus06} Dubus, G.\ 2006, \aap, 456, 801 


\bibitem[Dubus et al.(2008)]{Dubus2008} Dubus,~G., Cerutti, B., \& Henri, G. 2008, \aap, 477, 691

\bibitem[Fukazawa et al.(2009)]{fuka08} Fukazawa,~Y., et al. 2008, \pasj, 61, S17

\bibitem[Gruber et al.(1999)]{gruber99} Gruber,~D.~E., et al. 1999, \apj, 520, 124

\bibitem[Hillas(1984)]{Hillas} Hillas, A. M. 1984, \araa, 22, 425

\bibitem[Hoffmann et al.(2009)]{hoffmann} Hoffmann, A.~D., 
Klochkov, D., Santangelo, A., Horns, D., Segreto, A., Staubert, R., 
\& Puehlhofer, G.  2009, \aap, 494, L37 

\bibitem[Johnston et al.(1992)]{Johnston92} Johnston, S., et al.\ 1992, \apjl, 387, L37 

\bibitem[Khangulyan \& Aharonian(2005)]{mitya_felix05} Khangulyan, D., \& Aharonian, F. 2005, in AIP Conf. Proc., 745, 359

\bibitem[Khangulyan  et al. (2007)]{Khangulyan2007} Khangulyan, D., Hnatic, S., Aharonian, F., \& Bogovalov, S. 2007, 
\mnras, 380, 320

\bibitem[Khangulyan  et al. (2008a)]{Mitya_LS} Khangulyan, D, Aharonian, F., \& Bosch-Ramon, V. 2008a, 
\mnras,  383, 467

\bibitem[Khangulyan et al.(2008b)]{khangulyan_dublin} Khangulyan, D.~V., 
Aharonian, F.~A., Bogovalov, S.~V., Koldoba, A.~V.,\& Ustyugova, G.~V.\ 2008b, International Journal of Modern Physics D, 17, 1909

\bibitem[Kokubun et al.(2007)]{kok07} Kokubun, M., et al. 2007, \pasj, 59, SP1, 53

\bibitem[Koyama et al.(2007)]{koy07} Koyama, K., et al. 2007, \pasj, 59, SP1, 23

\bibitem[Krivonos et al.(2007)]{klivonos07} Krivonos,~R., et al. 2007, \aap, 463, 957

\bibitem[Marshall et al.(2002)]{Marshall02} Marshall, H.~L., 
Canizares, C.~R., \& Schulz, N.~S.\ 2002, \apj, 564, 941 

\bibitem[Martocchia et al.(2005)]{martocchia05} Martocchia,~A., Motch,~C., and Negueruela,~I. 2005, \aap, 430, 245

\bibitem[Mitsuda et al.(2007)]{mit07} Mitsuda, K., et al. 2007, \pasj, 59,  SP1, 1
\bibitem[Motch et al.(1997)]{Motch97} Motch, C., Haberl, F., Dennerl, K., Pakull, M., \& Janot-Pacheco, E.\ 1997, \aap, 323, 853 

\bibitem[Paredes et al.(2000)]{Paredes00} Paredes, J.~M., 
Mart{\'{\i}}, J., Rib{\'o}, M., \& Massi, M.\ 2000, Science, 288, 2340 

\bibitem[Paredes et al.(2002)]{Paredes02} Paredes, J.~M., Rib{\'o}, M., Ros, E., Mart{\'{\i}}, J., \& Massi, M.\ 2002, \aap, 393, L99 

\bibitem[Perucho \& Bosch-Ramon(2008)]{perucho08} Perucho, M. \& Bosch-Ramon, V. 2008, \aap, 482, 917

\bibitem[Rib$\acute{{\rm o}}$ et al.(1999)]{ribo99} Rib$\acute{{\rm o}}$,~M., Reig,~P., Marti,~J., and Paredes,~J.~M. 1999, A\&A, 347, 51

\bibitem[Serlemitsos et al.(2007)]{ser07} Serlemitsos, P. J., et al. 2007, \pasj, 59,SP1, 9

\bibitem[Torres \& Sierpowska-Bartosik(2008)]{diego08} Sierpowska-Bartosik \& Torres, D. 2008, Astroparticle Physics. Phys. 30, 239

\bibitem[Takahashi et al.(2007)]{tak07} Takahashi, T. et al. 2007, \pasj, 59, SP1, 35

\bibitem[Zabalza et al.(2008)]{zabalza08} Zabalza, V.,  Paredes, J. M., Bosch-Ramon, V., 2008, Int. Jour. Mod. Phys. D, 17, 1867

\end{thebibliography}
\end{document}